# Unveiling the effects of Cu doping on the H$_2$ activation by CeO$_2$ surface frustrated Lewis pairs

Tongtong Liu[a,†], Xinyi Wu[a,†], Kaisi Liu,*[a] and Lei Liu*[a]

[a] Center for Computational Chemistry, College of Chemistry and Chemical Engineering, Wuhan Textile University, Wuhan 430200, China

Corresponding authors:
Kaisi Liu, liuks@wtu.edu.cn
Lei Liu, liulei@wtu.edu.cn; liulei3039@gmail.com
[†] Equal contributions.

**Abstract:** Recently, the solid-state frustrated Lewis pairs (FLPs) on the surface of $CeO_2$ have been demonstrated to effectively catalyze the selective hydrogenation of unsaturated substrates, hence, the relationship between their intrinsic properties and $H_2$ activation at the atomic scale has attracted great attention. In this work, the effects of Cu doping on the intrinsic FLPs properties for different facets of $CeO_2$ is investigated by using density functional theory calculations, including the geometric parameters between Lewis acid-base centers, and the reactivity of Lewis acid-base towards $H_2$ activation. The study demonstrates that introducing O vacancies on different crystal facets of $CeO_2$ creates FLPs with the ability to efficiently cleavage hydrogen molecules. After the substitution of Ce with Cu, the inadequate electron availability of Cu to bond with O contributes to a reduction in the formation energy of O vacancies. Importantly, Cu exert an influence not only on the intrinsic properties of FLPs but also on the formation of new Ce-O and Cu-O FLPs. Considering the $H_2$ activation, the doping of Cu results in an enhancement for the thermodynamics by decreasing the reaction energies, while a hinderance for the kinetics by increasing the energy barriers. Overall, with these theoretical investigations, we propose certain hints for the future experimental studies concerning the synthesis of Cu doped $CeO_2$ catalysts for the $H_2$ activation and hydrogenation reactions.

**Keywords**: frustrated Lewis pairs; $CeO_2$; density functional theory; $H_2$ activation

**Introduction**

The activation of hydrogen plays a crucial role in the catalytic hydrogenation of unsaturated substrates, including alkenes,[1–3] alkynes,[4,5] and $CO_2$.[6,7] The activation of $H_2$ predominantly takes place on precious metals, including Ru,[8–10] Rh,[11] Pd,[12] Ir,[13] and Pt[14], which are often supported by carriers such as alumina,[15,16] silica,[17] or zeolites.[18] Although certain precious metals exhibit significant activity, their high cost or limited capability for low-temperature activation has constrained their further industrial development. Consequently, there has been significant recent interest in exploring alternative, economically viable catalysts for the activation of hydrogen.[19–22] One of the breakthrough studies has been reported by Stephan and co-workers, in which the authors proposed a concept of frustrated Lewis pairs (FLPs), based on the experiment observations of reversible $H_2$ activation by $Mes_2PCF_4B(C_6F_5)_2$ (Mes=2,4,6-$Me_3C_6H_2$).[23] Subsequently, FLP catalysts have shown significant reactivity in the activation of small molecules (e.g. $O_2$, $N_2$ and $CO_2$) and in the hydrogenation of unsaturated compounds (e.g. alkenes, alkynes, and carbonyl), becoming an emerging research topic.[24–30] Yet, homogeneous FLP catalysts based on molecular structures present challenges in catalyst recovery and product purification. Heterogeneous catalysts possess both high activity and ease of recovery advantages.[31] Therefore, the advancement of heterogeneous catalysts that FLPs constructed by a solid surface is highly anticipate. Ceria is one of the candidate heterogeneous catalysts for surface-constructed FLPs, owing to its different oxidation states ($Ce^{3+}/Ce^{4+}$ redox pair) and controllable surface structures. Zhang et al., initially constructed the solid FLPs on the $CeO_2$ (110) facet by introducing O vacancies to cleavage H-H bond with a low activation energy of 0.17 eV.[32] The reduced Ce serves as a Lewis acid site, while the neighboring lattice O on the surface functions as a Lewis base site. The distance between the Lewis acid site of Ce and Lewis base site of O is approximately 4 Å, significantly exceeding the typical Ce−O bond length (ca. 2 Å) observed in stoichiometric $CeO_2$, thereby conforming to the concept of FLPs. Moreover, Lewis acid sites of reduced Ce can stably adsorb dissociated hydride ($H^{\delta-}$), in contrast to the adsorption of $H^{\delta-}$ on the surface O sites of stoichiometric $CeO_2$, and the presence of O vacancies on the surface of $CeO_2$ can simultaneously enhance the strength of Lewis acid and Lewis base sites.[33]

Generally, the intrinsic properties of FLPs, such as the distance between Lewis acid-base sites and the strength of Lewis acid-base sites, are strongly related to the activity for $H_2$ activation.[34,35] Recently, metal ions in situ substitution of Ce represents an effective strategy for tuning the intrinsic properties of FLPs.[36–38] For example, the introduction of Ni onto the $CeO_2$ (110) facet influences the valence state of Ce.[39] Specifically, on $CeO_2$ (110) facet with an O vacancy (denoted as $CeO_2$ (110)-$O_v$), the Lewis acid sites of Ce primarily exhibit the +3 oxidation state. Upon doping $CeO_2$ (110)-$O_v$ with Ni, the Lewis acid sites of Ce exhibit a presence of the +4 oxidation state. Consequently, $CeO_2$ (110)-$O_v$ exhibits greater activity than Ni-doped $CeO_2$ (110)-$O_v$ due to Ce in the +3 oxidation state facilitates stronger binding ability with hydrogen molecules. For the $CeO_2$ (111) facet, doping Ni contributes to the formation of O vacancies. Ni doping does not directly participate in $H_2$ activation as FLPs, but functions as a single-atom promoter.[40] Upon doping $CeO_2$ (111)-$O_v$ with Ga, Ga not only function as single-atom promoters, but also serve as Lewis acidic sites of FLPs.[41] Cu-based catalysts have demonstrated favorable selectivity in both hydrogenation reactions and $CO_2$ reduction reactions.[42–44] Ban et al. identified that doping Cu on the $CeO_2$(110) facet can facilitate the generation of O vacancies, and enhance the formation of FLP sites.[45] Zhou et al. employed density functional theory (DFT) calculations to illustrate that Cu incorporation on the $CeO_2$ (111) facet generates Cu-O FLPs which promote the $H_2$ activation and $C_2H_2$ hydrogenation.[46] Overall, the intrinsic properties of FLPs on the surface of $CeO_2$ are intricately associated with the types of incorporation of metal ions and crystal facets.

Herein, we explored the intrinsic properties of FLPs and the activity for $H_2$ activation on (100), (110), and (111) crystal facets of $CeO_2$ with one O vacancy. Moreover, the impact of Cu doping on the intrinsic properties of FLPs and their reactivity with $H_2$ was investigated *via* DFT calculations. The study demonstrates that FLPs on (100), (110) and (111) facets containing one O vacancy are capable of efficiently activating $H_2$, and the formation energy of O vacancies for all studied facets decreases after the Cu doping. Moreover, the activity of FLPs has been changed as well due to the Cu doping, e.g. reaction energies of $H_2$ activation increase in the cases of (100), (110), and decrease in the case of (111), while the energy barriers increase in all studied cases.

**Computational details**

All computations were conducted by utilizing DFT theory as implemented in the

Vienna Ab initio Simulation Package (VASP).[47,48] The gradient-corrected Perdew–Burke–Ernzerhof (PBE) approximation was employed to address the exchange–correlation potential.[49] The wave functions of valence electrons were expanded using plane waves with a cutoff energy of 450 eV, whereas core electrons were modeled using the projector augmented-wave (PAW) method.[50] The van der Waals correction was included using the DFT-D3 method of Grimme.[51] The convergence criteria for forces acting on each iron atom and for energy were established at 0.02 eV/Å and $10^{-5}$ eV, respectively. The transition states of relevant elementary reaction steps were identified utilizing the nudged elastic band (NEB) method.[52] A periodic slab with p (3 × 4), p (3 × 5), p (4 × 4) unit cells was chosen to model the $CeO_2$ (100), (110) and (111) surfaces, respectively. To eliminate the dipole effects in the case of (100) crystal facet, we moved half O of one face to the opposite side, which was reported in a previous study.[53] A vacuum space of 15 Å was employed between the neighboring interleaved slabs. During all calculations, the atoms within the top three layers were fully optimized, whereas all the other atoms were fixed.

The O vacancy formation energy ($E_{f\text{-vac}}$) was defined as: $E_{f\text{-vac}} = E_{(\text{slab-}O_v)} + 1/2E_{(O_2)} - E_{(\text{slab})}$. The doping energy ($\Delta E_{\text{dope (Ce)}}$) was calculated by: $\Delta E_{\text{dope (Ce)}} = E_{Cu@CeO_2} - E_{CeO_2} - E_{Cu} + E_{Ce}$. The adsorption energy ($E_{\text{ads}}$) was computed using the equation: $E_{\text{ads}} = E_{(\text{adsorbate + surface})} - E_{(\text{free molecule})} - E_{(\text{free surface})}$.

**Results and Discussion**

**Structures of (100) facet.** For the perfect $CeO_2$ (100) facet, the distances between unbonded Ce and O (e.g. CeII-OIII, and CeII-OII in Fig. 1a) are 4.43 Å and 4.41 Å, respectively, potentially exhibiting FLP-like catalysts based on the geometric parameters, which has been identified to be from 3 Å to 5 Å.[35,54] However, the electronic interactions between CeII and its adjacent OI atom impedes the reactivity of CeII-OIII and CeII-OII (e.g. with a positive computed reaction energy for $H_2$ activation, see $H_2$ activation section). Upon removal of the OI atom to introduce an O vacancy (Fig. 1b, denoted as (100)-$O_v$), the CeII and surface lattice oxygen OII and OIII become Lewis acid and base active centers. After geometry relaxation, the distances between Ce and O in CeII-OIII and CeII-OII pairs of (100)-$O_v$ are measured to be 4.70 Å and 4.40 Å, respectively, due to the structural distortion. These values of distance indicate that CeII-OIII and CeII-OII pairs might serve as FLP catalysts (denoted as FLPs (100)-$O_v$-I and (100)-$O_v$-II in Fig. 1b). In the case of (100)-$O_v$, the Bader charges of surface

OIII and OII are -1.18 $e$ and -1.12 $e$, respectively (Table S1). This reflects a respective increase of +0.01 $e$ and +0.04 $e$ in the comparison to their corresponding values of the perfect (100) surface, indicating an enhancement in the basicity of OII and OIII. Moreover, upon the introduction of an O vacancy, the Bader charge of the corresponding CeII diminishes from +2.23 $e$ to +2.18 $e$. Subsequently, we substituted CeI with Cu to explore the effects of Cu doping on the structure and reactivity of FLPs (Fig. 1c). Owing to the valence electron configuration of Cu atom is 3d10 4s1, it lacks the ability to engage in hexacoordination with oxygen atoms. The distance between the Ce and O in the FLP of CeII-OIII decreases from 4.70 Å to 4.56 Å, and from 4.40 Å to 4.33 Å. The Bader charge of OIII increases from -1.18 $e$ to -1.19 $e$ (Table S1), while the Bader charge of OII decreases from -1.12 $e$ to -1.11 $e$ (Fig. 1f), and the acidity of the reduced CeII remains unchanged. Interestingly, a new FLP was found after the Cu doping between CeIII and OIV, denoted as Cu@CeO$_2$ (100)-O$_v$-III, as shown in Fig. S1. This FLP forms as follows: after substituting of CeI with Cu, OV migrates to locate at between CeIII and CeIV during the geometry optimization (see Fig. S2). Bader charge of this FLP was computed to be +2.18 $e$ and -1.09 $e$ for CeIII and OIV, respectively.

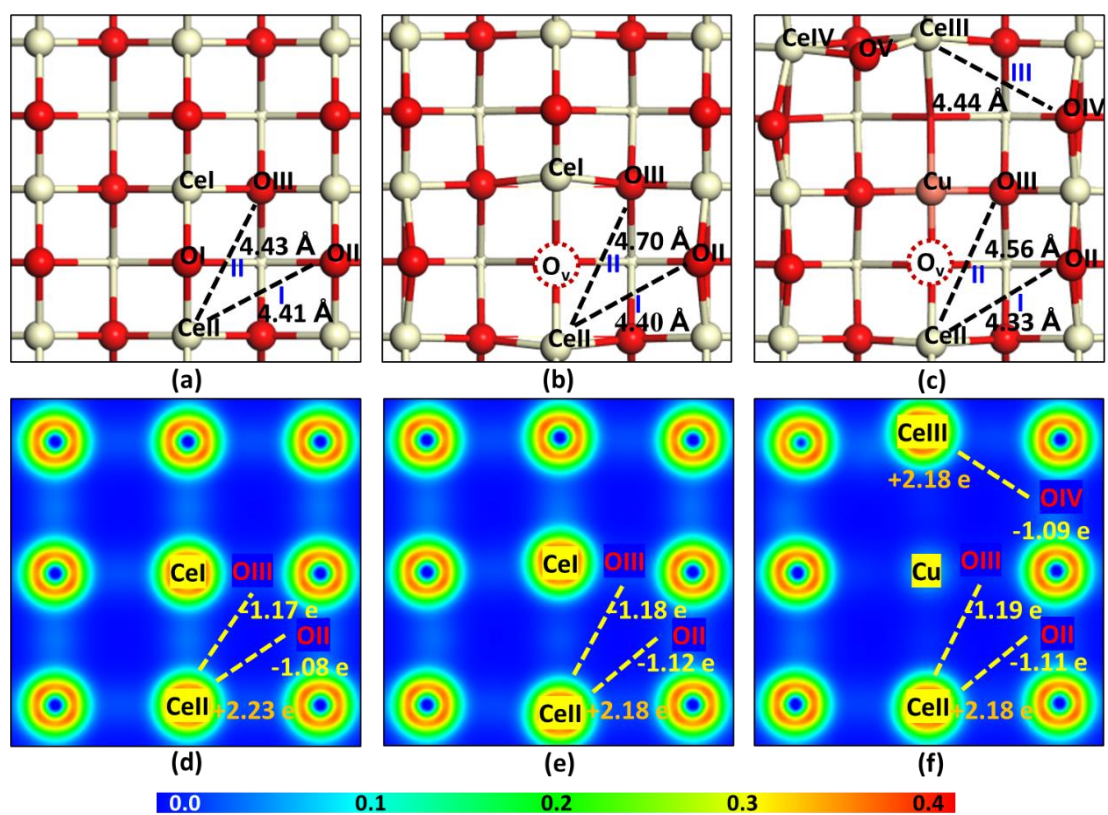

**Figure 1.** Optimized structure of perfect $CeO_2$ (100) (a), $CeO_2$ (100)-$O_v$ (b), and $Cu@CeO_2$ (100)-$O_v$ (c). Charge-density isosurfaces of perfect $CeO_2$ (100) (d), $CeO_2$ (100)-$O_v$ (e), and $Cu@CeO_2$ (100)-$O_v$ (f). The charge-density isosurfaces are plotted at 0.05 $e$ bohr$^{-3}$. Color legend: Cu, brownish red; Ce, faint yellow; O, red.

**Structures of (110) facet.** Analogous to the (100) facet, the surface FLPs can also be formed by selectively removing surface O from the (110) surface. In general, the situation of (110) is somehow simpler compare to that of (100) facet, of which only one FLP has been found after introducing of O vacancy and doping of Cu. As shown in Fig. 2a, the distance between unbonded CeII and OIII in the perfect (110) facet was computed to be 4.56 Å. Similar to that of (100) facet, this Lewis acid and base pair is again unable to active $H_2$ with a computed reaction being 0.3 eV. When the OI is removed, that is introducing of O vacancy, a FLP was formed between CeII and OIII, with distance to be 4.49 Å (denoted as FLP (110)-$O_v$-I in Fig. 2b). In contrast to the (100) facet, the (110) facet exhibits a decrease in the basicity of OIII, which has a smaller negative charge of -1.10 $e$, and an increase in acidity of CeII, which has a large positive charge of +2.29 $e$ (Fig. 2e and Table S1). To gain deeper into the impact of Cu, CeII is then replaced by a Cu. Due to the Cu-O tetrahedral coordination configuration, the position of OII has migrated, which is shared by two Ce from two layers after the geometry optimization (see Fig. S2), thereby leading to an increase in the distance between CeII and OIII from 4.49 Å to 4.62 Å. According to the Bader charge analysis, the charge of the Lewis acid CeII decreases from +2.29 $e$ to +2.26 $e$, whereas the Bader charge of the Lewis base OIII increases from -1.10 $e$ to -1.15 $e$, indicating that the introduction of Cu doping in the (110) facet results in a reduction in the acidity and an increase in the basicity of FLPs.

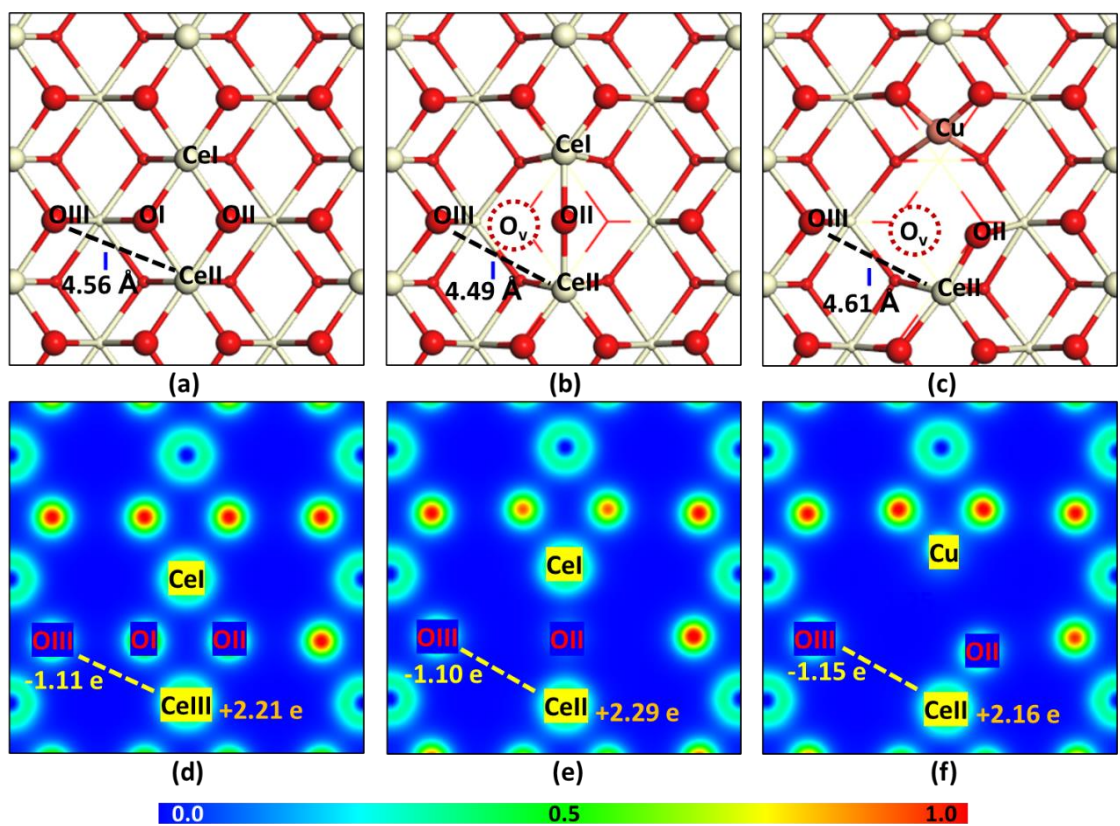

**Figure 2.** Optimized structure of perfect $CeO_2$ (110) (a), $CeO_2$ (110)-$O_v$ (b), and Cu@$CeO_2$ (110)-$O_v$ (c). Charge-density isosurfaces of perfect $CeO_2$ (110) (d), $CeO_2$ (110)-$O_v$ (e), and Cu@$CeO_2$ (110)-$O_v$ (f). The charge-density isosurfaces are plotted at 0.05 $e$ bohr$^{-3}$. Color legend: Cu, brownish red; Ce, faint yellow; O, red.

**Structures of (111) facet.** In the case of (111) facet three surface FLPs were found upon introducing O vacancy and doping Cu individually. In details, removing OI on the (111) facet results in the formation of two types FLPs, denoted as (111)-$O_v$-I, which contains CeII-OII, (111)-$O_v$-II, which contains CeIII-OIII, and (111)-$O_v$-III, which contains CeII-OIV (see Fig. 3b). The Ce and O in first FLP (CeII-OII) located at the same layer (e.g. the top layer) and their distance was computed to be 4.69 Å, which is larger than that in the perfect (111) facet, being 4.50 Å as shown in Fig. 3a. The Ce and O in last two FLPs located at different layers, e.g. Ce at the top layer and O at the second layer. The distance between Ce and O of these two FLPs are shorter compared to the first one, being 4.45 Å. Upon replacing CeI with Cu, the Cu-O tetrahedral coordination prevents the formation of a covalent bond between Cu and OII, and introduces locally structural distortion (Fig. 3c). As such, the distance between OII and CeII increases to be 4.80 Å. Moreover, Bader charge analysis reveals that the Lewis acidity of CeII and CeIII almost remain the same with charge changes less than 0.02 e, while Lewis basicity of OII, OIII and OIV decreases with charges decrease from ca. +1.17 $e$ to ca. +1.07 $e$ (Table S1).

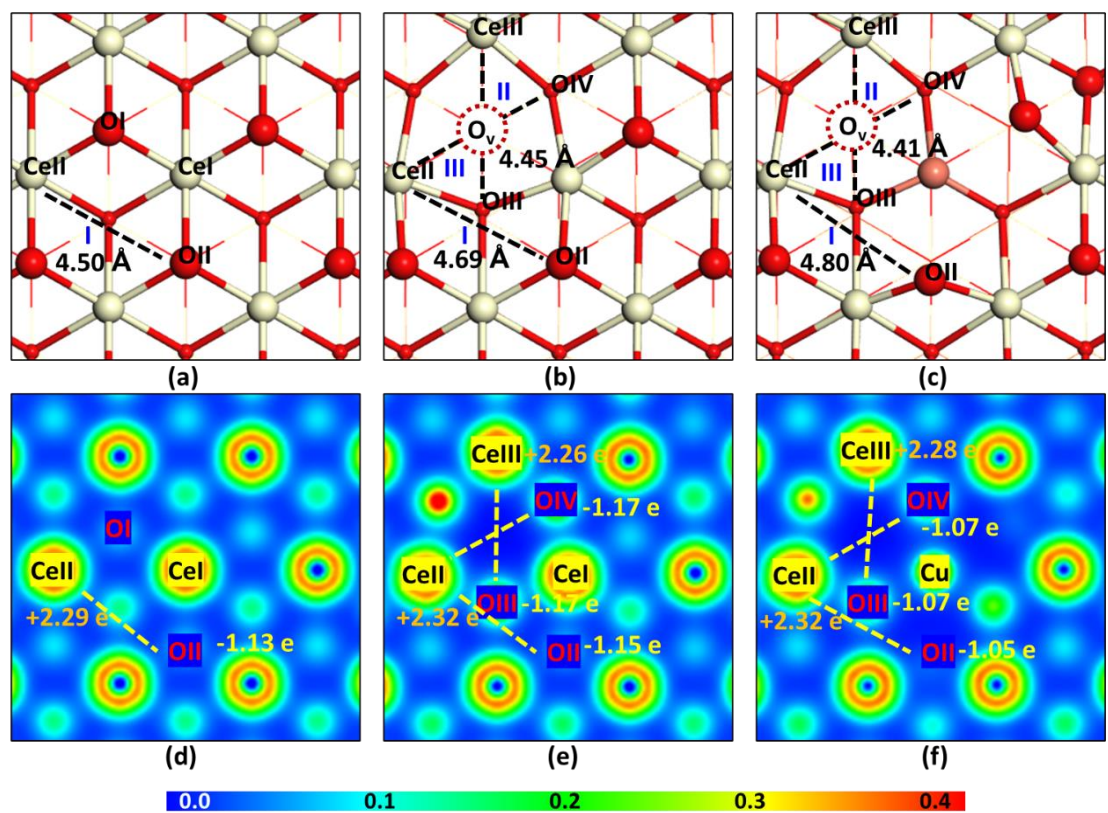

**Figure 3.** Optimized structure of perfect $CeO_2$ (111) (a), $CeO_2$ (111)-$O_v$ (b), Cu@$CeO_2$ (111)-$O_v$ (c). Charge-density isosurfaces of perfect $CeO_2$ (111) (d), $CeO_2$ (111)-$O_v$ (e), Cu@$CeO_2$ (111)-$O_v$ (f). The charge-density isosurfaces are plotted at 0.05 $e$ bohr$^{-3}$. Color legend: Cu, brownish red; Ce, faint yellow; O, red.

In short, surface FLPs could be constructed on the crystal facets of $CeO_2$ (100), (110), and (111), through the introduction of oxygen vacancies, and the electron-donating ability of Lewis bases is often enhanced based on the Bader charge analysis. Upon Cu doping, obvious structural distortions have been obtained since Cu only bond with four O instead of six in the case of Ce. Importantly, we found that the influence of Cu on the acidity and basicity of Lewis pair and base varies with different crystal facets. Moreover, Cu doping might also lead to the formation of new Cu-O FLPs according to the distance criteria of FLPs. For example, Cu-OII with distance of 4.80 Å on the (100) facet, Cu-OII with distance of 3.65 Å on the $CeO_2$ (110) facet, and Cu-OIV with distance of 4.23 Å on the $CeO_2$ (111) facet (Fig. S1).

**Formation energies.** As shown in the Fig. 4a, the formation energies of one O vacancy on perfect $CeO_2$ (100), (110), and (111) crystal facets are 2.81 eV, 2.22 eV and 3.30 eV, respectively (corresponding optimized structures are shown in Fig. S3). Note that, the formation energy of creating one O vacancy on the (110) facet is the lowest, which is quite consistent with a previous study,[32] indicating that the model and the method employed in our work is reliable. Interesting, we found that after doping Cu into crystal facets, the formation energies of creating one O vacancy are showed notable decrease to be -0.95 eV, 0.13 eV, and -0.44 eV, respectively for (100), (110) and (111). These observations might be attributed to Cu atom lacks sufficient electrons to form covalent bonds with the corresponding O atoms. As discussed in the structure section, Ce usually has six coordination with O, while Cu coordinates with only four O atoms. It is worth to point out that, after Cu doping, the formation energy of oxygen vacancies on the (110) facet becomes the highest compared to that of (100) and (111) facets, while a perfect (110) has the lowest formation energy of one O vacancy. We then compute the doping energies of Cu for systems of Cu@CeO$_2$ and Cu@CeO$_2$-O$_v$, and the values are summarized in Fig. 4b. In general, the doping energies for CeO$_2$ facets containing one O vacancy are notably reduced compared to that of perfect CeO$_2$ facets, indicating the introduce of O vacancy on CeO$_2$ surface facilitates the doping of Cu. In details, the doping energy of Cu on the (111) facet is the highest, which may be attributed to substituted Ce atom is seven O coordinated, in contrast to the six O coordinated environment of the substituted Ce atoms in the (100) and (110) crystal facets (see Fig. S4). As a conclusion, the incorporation of Cu effectively reduces the formation energy of oxygen vacancies, thereby facilitating the formation of FLPs.

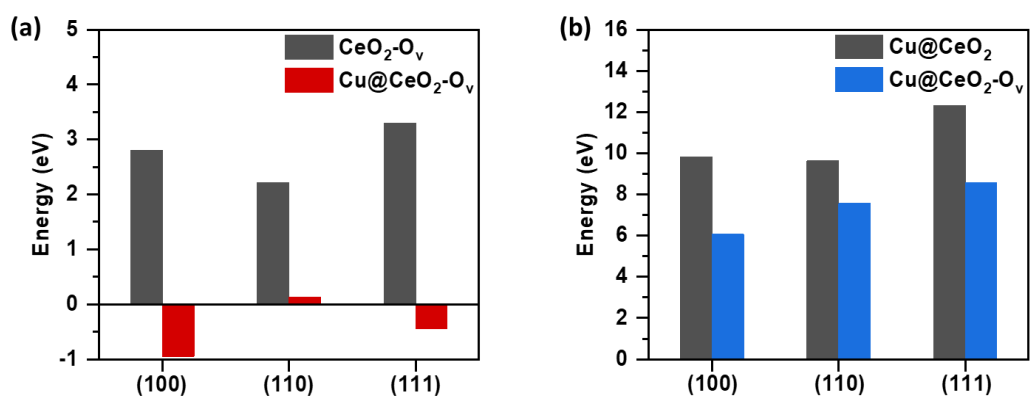

**Figure 4.** The formation energies of creating one O vacancy on the different facets of perfect CeO$_2$ (black) and Cu@CeO$_2$ (red). The doping energies of Cu on the different facets of perfect CeO$_2$ (black) and CeO$_2$-O$_v$ (blue).

**Hydrogen activation.** Lastly, we computed the reaction energies and energy barriers for the $H_2$ activation by above-mentioned FLPs, and the corresponding profiles are given in Fig. 5. For (100) facet with one O vacancy, the reaction energies for the $H_2$ activation by two FLPs of CeII-OIII and CeII-OII are -0.63 eV and 1.12 eV, respectively, which are 1.69 eV and 0.54 eV lower than the reaction energies by the configurations of CeII-OIII and CeII-OII on perfect (100) facet (Fig. 5a left column). This indicates that FLPs formed after the introduction of an O vacancy can effectively promote the $H_2$ activation, while the perfect facets are not able to cleave the H-H bond in the $H_2$ molecule. After doping Cu onto (100)-$O_v$ facet, the reaction energies for $H_2$ activation by CeII-OIII and CeII-OII FLPs are -0.47 eV and 0.66 eV, respectively, which illustrates that Cu doping on the (100) facet reduces the activity of FLPs for $H_2$ activation. Moreover, the new CeIII-OIV FLP formed due to the Cu doping also could effectively perform the $H_2$ activation, with a computed reaction energy being -0.26 eV. Subsequently, we investigated the reaction kinetics of $H_2$ activation by FLPs which showed negative reaction energies. The adsorption energies of $H_2$ on (100)-$O_v$ and Cu@$CeO_2$ (100)-$O_v$ facets are comparable, which are ca. -0.23 eV, indicating that adsorption of $H_2$ on these surfaces are dominated by week van der Waals interactions, and Cu doping shows negligible impacts. The energy barrier for H-H bond cleavage by CeII-OII on (100)-$O_v$ facets is 0.08 eV, which is lower than that of the Cu doping cases, e.g., CeII-OII and CeIII-OIV on Cu@$CeO_2$ (100)-$O_v$ facet, whose energy barriers are 0.16 eV and 0.30 eV, respectively. In short, Cu doping slightly increases the energy barriers for $H_2$ activation on (100) facets. Yet, all studied cases could be able to cleavage the H-H bonds in term of kinetics. For the (110) facet, all FLPs constructing by CeII-OIII show relatively low reaction energies, varying from -0.05 eV to 0.32 eV for the $H_2$ activation (Fig. 5a middle column). We conclude two interesting points: 1) the perfect (110) facet (that is without O vacancies and Cu doping) might also work for $H_2$ activation, of which the computed reaction energy is a slight positive value of 0.32 eV. This observation is quite different from the case of (100) and (111) facets, where the computed reaction energies are rather positive, being 1.12 eV and 1.06 eV, respectively; 2) talking about further hydrogenation reactions, (110) would be a better choice, since utilization of $H_2$ resource requires a zero or close to zero value for $H_2$ activation reaction energy, which is not the case for all FLPs on the (100) and (111) facets, since some of the FLPs show relative large negative values of reaction energies for $H_2$ activation. Moreover, we also found that doping Cu increase the energy barrier, e.g., from 0.39 eV to 0.64 eV. For (111) facet, a similar trend has been found, that is, the perfect facets show a positive value for the reaction energy, and introducing O vacancy decrease the reaction energy, while Cu doping decreases further to be negative for all FLPs (Fig. 5a right column). In case of reaction path profile show in Fig. 5b right column, we observe that the $H_2$ adsorption energy of CeII-OII FLP decreases from -0.17 eV to -1.06 eV, demonstrating Cu doping enhances the property of FLPs to adsorb $H_2$. Yet, the energy barrier for such FLP to cleave the H-H bond increase from 0.51 eV to 1.39 eV, indicating that Cu doping might hinder the $H_2$ activation by CeII-OII FLP. For CeII-OIV and CeIII-OIII pairs, we also see Cu doping strengths the $H_2$ adsorption and increase the energy barriers for $H_2$ activation.

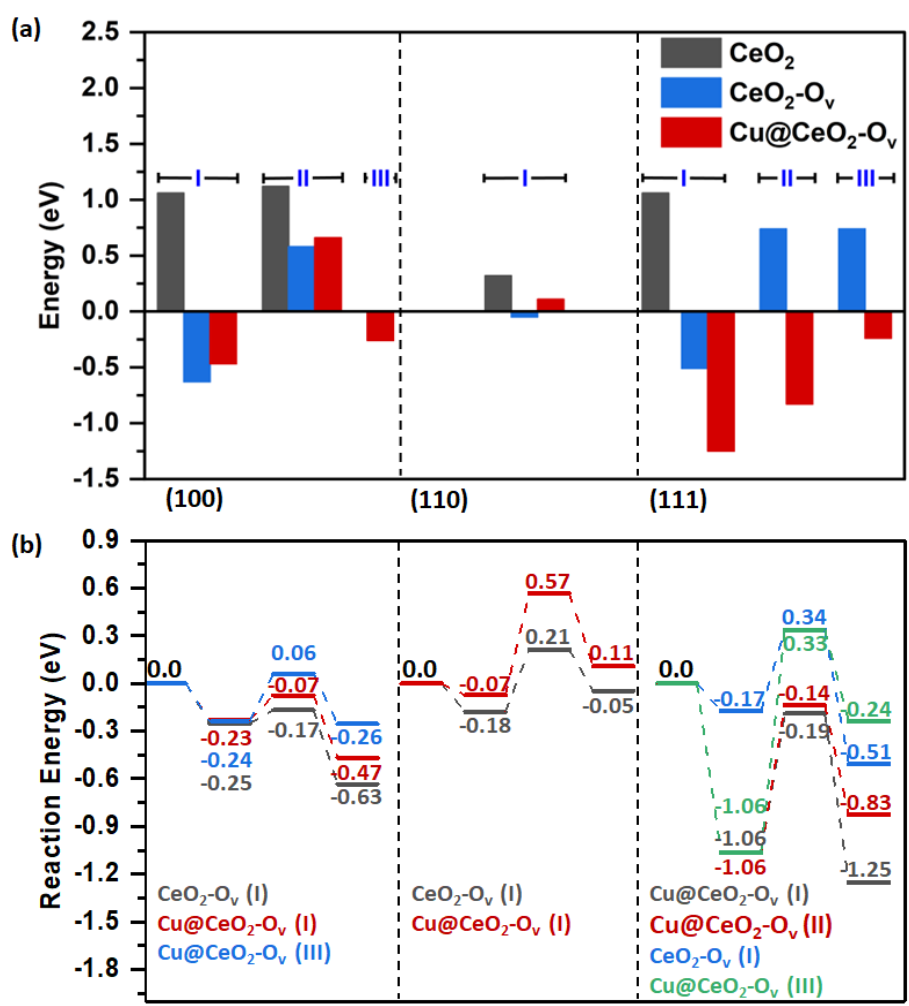

**Figure 5.** Reaction energies of $H_2$ activation (a), and the energy profiles of $H_2$ activation pathways on the different facets of $CeO_2$-$O_v$ with the notation defined in **Fig. 1** to **Fig. 3**.

As a summary, Cu doping influences the distance and strength of Lewis acid-base centers in all Ce-O FLPs, and thereby affecting the ability of FLPs to activate $H_2$.

Generally, Cu doping enhances the thermochemistry for $H_2$ activation by decreasing the reaction energies, while introduces certain issues for kinetics by increasing the energy barriers. Moreover, we also examined the reaction energies for $H_2$ activation by Cu-O FLPs, which are large negative values being -6.03 eV, -4.00 eV and -2.95 eV, respectively (see Fig. S5). This is because that following geometry optimization, the H(Cu) atoms transformed onto O atoms forming more stable structures. These observations are analogous to the $H_2$ activation by Ni-O FLPs on the $CeO_2$ (110) facet.[40] We should note that such large negative values are somehow not suitable for post utilization of $H_2$ resource, therefore, further studies should be performed to avoid such situations.

**Conclusions**

In this work, DFT calculations were conducted to investigate the $H_2$ activation by FLPs on the perfect (100), (110) and (111) facets of $CeO_2$, and with O vacancy as well as Cu doping. The results show that the introduction of O vacancies on the surface successfully creates FLPs that catalyze $H_2$ activation. Moreover, as Cu can only form a tetrahedral coordination with O, Cu doping promotes the formation of O vacancies by decreasing the formation energy of O vacancies, and even leads to the generation of new Ce-O and Cu-O FLPs by the distortion of the structures. Interestingly, with Cu doping, we found that reaction energies for $H_2$ activation by FLPs on the $CeO_2$ (100) and (110) crystal facets in some situation increases, whereas the reaction energies decrease on the $CeO_2$ (111) facet in all studied cases. Moreover, we found that Cu doping always increase the energy barriers. Hence, the balance for the thermodynamics and kinetics should be carefully considered upon Cu doing in the future experimental investigates. Another issues should be also taking care for Cu doping is the formation of Cu-O FLPs, of which large negative values were found for the reaction energies, which might hinder the post utilization of $H_2$, e.g., hydrogen reduction of $CO_2$ or unsaturated compounds. Overall, this study elucidates in detail the impact of Cu doping on the intrinsic properties of FLPs on (100), (110) and (111) facets of $CeO_2$, paving the way for subsequent exploration of Cu-doped $CeO_2$ (Cu@$CeO_2$) catalysts in the hydrogenation of unsaturated alkenes, alkynes, and $CO_2$.

**Acknowledgement**

L.L is thankful for the financial support from the National Natural Science Foundation (No. 21978294), and start-up funding from Wuhan Textile University (No. 20220321).

## Conflicts of Interest

The authors declare that they have no known competing financial interests.